\documentclass[aps,pra,floatfix,footinbib,preprintnumbers,superscriptaddress,notitlepage,reprint]{revtex4-2}

\usepackage{amsmath,amssymb,mathtools}
\usepackage{bbold}
\usepackage{graphicx}
\usepackage{bm,color}
\usepackage{natbib,hyperref}
\usepackage[capitalise]{cleveref}
\usepackage{enumitem}
\usepackage{adjustbox}
\usepackage{mathrsfs}
\usepackage[normalem]{ulem}
\usepackage{xspace}

\hypersetup{breaklinks=true,colorlinks=true,linkcolor=blue,urlcolor=blue,citecolor=blue}

\newcommand{\be}{\begin{equation}}
\newcommand{\ee}{\end{equation}}

\newcommand{\sbra}[1]{\langle #1 |}
\newcommand{\sket}[1]{| #1 \rangle}
\newcommand{\ketbra}[2]{\ensuremath{|#1 \vphantom{#2} \rangle \langle #2 \vphantom{#1} | }}

\newcommand{\saj}[2]{\sigma^{#1}_{#2}}
\newcommand{\thetaopt}{\vec{\theta}_o}

\begin{document}
\preprint{FERMILAB-PUB-24-0103-ETD}
\title{Noise-induced transition in optimal solutions of variational quantum algorithms}
\author{Andy C.~Y.~Li}
\affiliation{Fermi National Accelerator Laboratory, Batavia, IL 60510}
\author{Imanol Hernandez}
\affiliation{Fermi National Accelerator Laboratory, Batavia, IL 60510}
\affiliation{Department of Physics and Astronomy, University of California Los Angeles, Los Angeles, CA}
\date{\today}

\begin{abstract}
Variational quantum algorithms are promising candidates for delivering practical quantum advantage on noisy intermediate-scale quantum (NISQ) hardware. However, optimizing the noisy cost functions associated with these algorithms is challenging for system sizes relevant to quantum advantage. In this work, we investigate the effect of noise on optimization by studying a variational quantum eigensolver (VQE) algorithm calculating the ground state of a spin chain model, and we observe an abrupt transition induced by noise to the optimal solutions. We will present numerical simulations, a demonstration using an IBM quantum processor unit (QPU), and a theoretical analysis indicating the origin of this transition. Our findings suggest that careful analysis is crucial to avoid misinterpreting the noise-induced features as genuine algorithm results.
\end{abstract}

\maketitle

\section{Introduction}
Variational quantum algorithms are considered a promising class of algorithms that can potentially demonstrate a practical quantum advantage in problems of scientific interest on NISQ hardware \cite{Preskill2018,Bharti2022}. This immense potential originates from the shallow variational circuits enabled by the hybrid nature with the optimization task offloaded to a classical computer and the fact that the optimization can mitigate certain coherent noises by shifting the cost function landscape according to the noises \cite{Peruzzo2014,mcclean2016,O'Malley2016,Bharti2022}. These make the variational algorithms have a much higher error tolerance than other quantum algorithms. In recent years, there have been many successful demonstrations, for example, of quantum machine learning \cite{Schuld2019,Havlicek2019,Peters2021} and with problems in quantum chemistry \cite{Peruzzo2014,O'Malley2016,Kandala2017} and condensed matter physics \cite{Kandala2017,Google2020,Zhang2022,Li2023} using NISQ hardware without error correction.

Implementing variational quantum algorithms on problems beyond the classical limit remains challenging due to the complexity of the nonconvex optimization performed on the classical side associated with the algorithms \cite{Cerezo2023}. One of the challenges is the barren plateau, where the cost-function gradient exponentially diminishes with an increase in the system size \cite{McClean2018,Arrasmith2021,Anschuetz2022}. Local minima appear in the high-dimensional cost-function landscape, making it difficult for local optimization to obtain a good solution \cite{Wiersema2020,Lee2021,Larocca2023}. Furthermore, analyzing the optimization complexity is made even more complicated by the noisy aspect resulting from the non-negligible error rates of NISQ hardware \cite{Wang2021,Wang2021,Stilck2021}.

In this manuscript, we studied the effect of noise on the optimal solution of a VQE algorithm calculating the ground state of a one-dimensional spin chain. We discovered that the noise can trigger a transition in the optimal solution. This transition can be qualitatively explained by the first-order perturbative correction to the cost-function landscape, which causes the global minimum to switch from one local minimum to another illustrated by the schematic diagram in \cref{fig:schematic}(a). We also demonstrated this transition on an IBM QPU for a spin dimer.
The noise-induced transition, which may occur for some applications of variational quantum algorithms, can result in the deterioration of the solution's quality in an unpredictable manner. Our study provides insights into the origin of this issue and paves the way for further studies in mitigating it to ensure the proper functioning of variational algorithms on NISQ hardware.

\section{Noisy VQE for XXX antiferromagnetic model}

\begin{figure*}
	\centering
	\includegraphics[width=\linewidth]{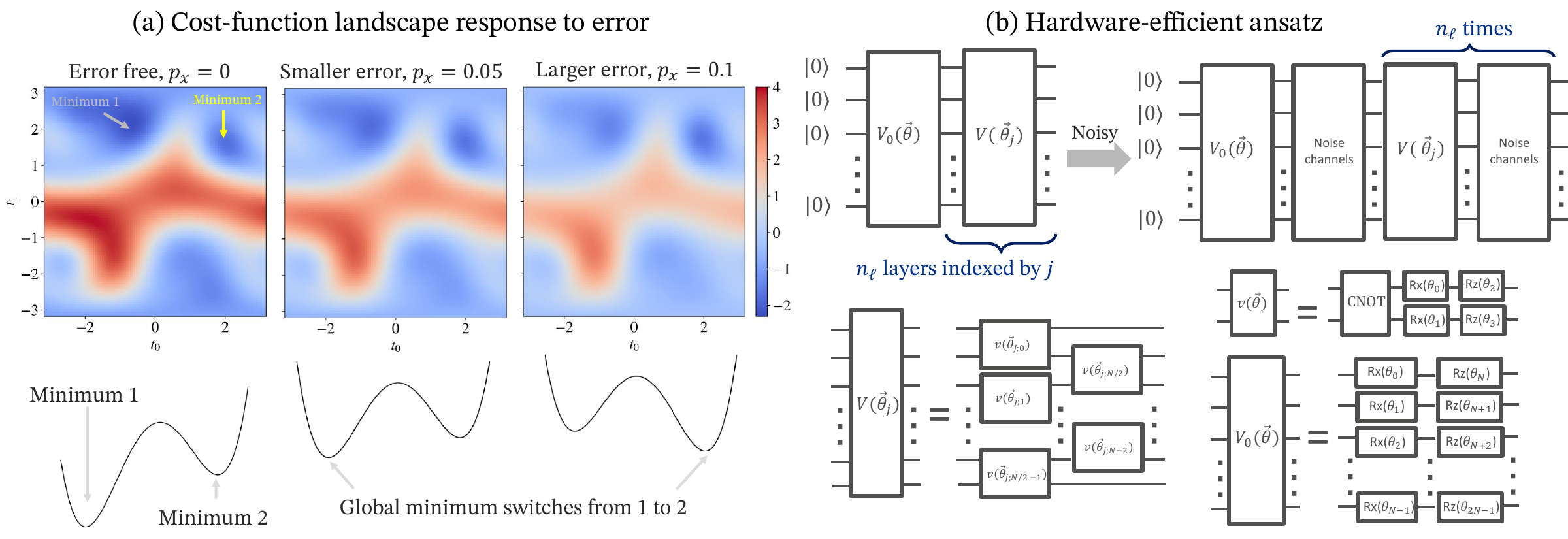}
	\caption{\textbf{Cost-function response to error and the associated hardware-efficient ansatz}.
		(a) We visualize the cost-function (energy) landscape corresponding to the two-parameter ansatz for the spin dimer. Two minima, labeled minimum 1 and minimum 2, can be found. In the error-free case with the bit-flip error rate $p_x = 0$, minimum 1 is the global minimum. With increasing error rate $p_x$, the energy associated with minimum 1 increases  faster than that with minimum 2. Beyond a threshold value, the global minimum switches from minimum 1 to minimum 2, and this results in a transition in the optimal solution of the variational algorithm.
		(b) We consider a hardware-efficient ansatz with $n_{\ell}$ layers. Within each layer, the qubits are entangled by CNOT gates, which are native to the IBM QPUs, and the CNOT gates are arranged in this particular structure such that they can be easily mapped to a linear qubit connectivity. Two parameterized single-qubit gates, Rx (rotation along the x-axis) and Rz (rotation along the z-axis), follow each CNOT gate. Our VQE algorithm optimizes all these rotation angles to approximate the true ground state.
	} 
	\label{fig:schematic}
\end{figure*}

We study the VQE optimal solution of a one-dimensional Heisenberg XXX model with antiferromagnetic coupling \cite{Heisenberg1928}.
The Hamiltonian of the spin chain with open boundary conditions is given by 
\be
\label{eq:model}
H = J \sum_{\alpha=X, Y, Z} \sum_{j=1}^{N - 1} \saj{\alpha}{j}\saj{\alpha}{j + 1} + h \sum_{\alpha=X, Y, Z} \sum_{j=1}^{N} \saj{\alpha}{j},
\ee
where $\saj{\alpha=X, Y, Z}{j}$ are the Pauli matrices on site $j$, $N$ is the number of spins, $J>0$ is the antiferromagnetic coupling strength, and $h$ is a magnetic field uniformly applied on X, Y, and Z directions.
The XXX antiferromagnetic spin chain has been widely studied with rich entanglement features \cite{Avdeev1986,Korepin1994,Jin2004,Popp2005}. Nonetheless, we merely use this model as a test bed to investigate the response of the VQE solutions to noise. We pick $J = h = 1$ for this purpose throughout this work.

VQE algorithms determine the ground state of the model by minimizing the cost function, the energy expectation value $E(\vec{\theta}) = \sbra{\psi(\vec{\theta})} H \sket{\psi(\vec{\theta})}$, associated with a parameterized ansatz $\sket{\psi(\vec{\theta})}$, where $\vec{\theta}$ is the parameters to be optimized \cite{Peruzzo2014,mcclean2016}. The optimal solution $\thetaopt$ gives the approximated ground state $\sket{\phi_0}$ and ground-state energy $E_0$ such that
\be
\thetaopt = \underset{\vec{\theta} }{\mathrm{argmin}} E(\vec{\theta}), \text{\ }
\sket{\phi_0} = \sket{\psi(\thetaopt)} \text{\ and \ }
E_0 = E(\thetaopt) 
.
	\ee
The quality of the solution representing the true ground state $\sket{\phi_g}$ and ground-state energy $E_{g}$ depends on the ansatz and the optimizer. Various ansatzes have been proposed with popular choices such as Hamiltonian variational ansatz \cite{Wecker2015} and hardware-efficient ansatz \cite{Kandala2017}. In this work, we use a hardware-efficient ansatz shown in \cref{fig:schematic}(b) with CNOT gates native to IBM QPUs to facilitate the hardware demonstration.

The VQE calculation is subject to coherent and incoherent errors running on NISQ hardware \cite{Kjaergaard2020,DeLeon2021}. Precise modeling these errors on a specific hardware is often challenging. Nonetheless, it is a reasonably good approximation to study the noise effect qualitatively with the depolarizing channel \cite{nielsen2002quantum}. For theoretical analysis and numerical simulation, we consider the bit-flip error for each qubit such that the density matrix $\rho$ representing the QPU quantum state evolves as
\be
\rho \rightarrow \rho' = (1 - N p_x) \rho + p_x \sum_{j=1}^{N}\saj{X}{j} \rho \saj{X}{j},
\ee
where $p_x$ is bit-flip error rate. We assume that bit-flip error with a rate $p_x$ occurs after each ansatz layer as shown in \cref{fig:schematic}(b).
The noisy cost function is given by $E(\vec{\theta}) = \mathrm{Tr}(\rho(\vec{\theta}) H)$ where $\rho(\vec{\theta})$ is the density matrix corresponding to the variational ansatz in the presence of errors. Note that for the error-free case, $\rho(\vec{\theta}) = \ketbra{\psi(\vec{\theta})}{\psi(\vec{\theta})}$ is a pure state and the cost function is equivalent to the above definition.

\begin{figure*}
	\centering
	\includegraphics[width=\linewidth]{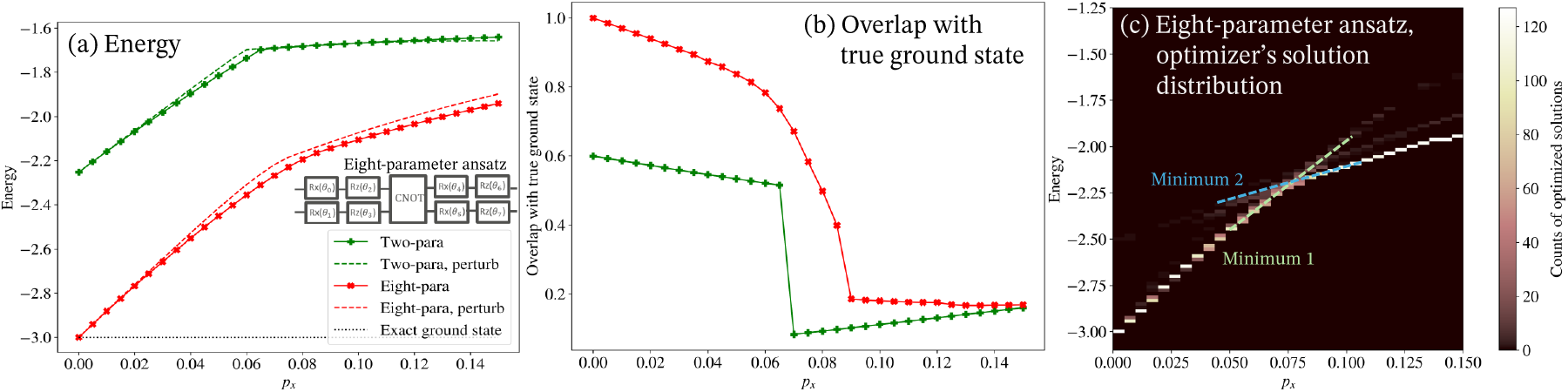}
	\caption{\textbf{Spin dimer noisy VQE results}.
		(a) The energy of the optimal solution with the eight-parameter ansatz (red crosses) shows a transition at around $p_x=0.07$. While the result from the two-parameter ansatz (green pluses) is not as good as the eight-parameter one in approximating the true ground state (dotted line), it exhibits a similar transition behavior.
		The optimal solutions obtained from the approximated noisy cost function, keeping only the leading-order error terms (dashed lines), are consistent with the ones obtained by the exact cost function (solid lines).
		The inset shows the circuit of the eight-parameter ansatz, and the two-parameter ansatz is defined by setting $\theta_0 = \theta_1 = -\theta_4=-\theta_5 = t_0$ and $\theta_2 = \theta_3 = -\theta_6=-\theta_7 = t_1$.
		(b) Noise-induced transitions can also be observed in the overlap between the optimal solution and the true ground state.
		(c) The local or global minimum energies can be inferred from the optimizer's solution distribution. We can observe the crossover between minima 1 and 2, which indicates the transition.
	} 
	\label{fig:dimer_result}
\end{figure*}

\section{Spin dimer simulation}

We start with the dimer case with two spins and a one-layer ansatz shown in the inset of \cref{fig:dimer_result}(a) consisting of eight parameters $\{\theta_k \}_{k=0, 1, 2, \cdots, 7}$. To facilitate analysis, we also consider a further simplified two-parameter ansatz by setting $\theta_0 = \theta_1 = -\theta_4=-\theta_5 = t_0$ and $\theta_2 = \theta_3 = -\theta_6=-\theta_7 = t_1$. 
Even though the simplified two-parameter ansatz is not as good as the eight-parameter ansatz in approximating the true ground state, it still preserves the essential characteristics qualitatively under increasing error rate. As a result, it enables us to gain a better understanding of the noise-induced transition by visualizing the cost-function landscape.

We optimize the cost functions associated with the two ansatzes with the \emph{SciPy} \cite{2020SciPy-NMeth} implementation of the L-BFGS-B optimization algorithm \cite{Liu1989}, which is a gradient-based local optimizer. To address the issue of local minima, we run the optimizer multiple times with 128 different initial parameters randomly chosen in the range of $[-\pi, \pi]$. We select the optimal parameters $\thetaopt$ as the solution associated with the lowest energy among the 128 runs.
In this study, we use the density-matrix formalism to simulate the noisy cost function and do not take sampling error into account.

The optimal-solution energies as a function of error rate $p_x$ are shown in \cref{fig:dimer_result}(a). The eight-parameter ansatz (red crosses) performs better than the two-parameter ansatz (green pluses) as the energy determined by the eight-parameter ansatz is nearly the same as the exact ground-state energy (black dotted line) in the zero-error case ($p_x = 0$). Nonetheless, the two share similar behavior with increasing $p_x$ in that the energy increases linearly with $p_x$ before seeing an abrupt transition at around $p_x = 0.07$ to a linear energy increase with a different slope.

The overlaps with the true ground state exhibit a more discontinuous transition as shown in \cref{fig:dimer_result}(b). Here, the overlap $f = \sbra{\phi_g} \rho(\thetaopt) \sket{\phi_g}$ between the ansatz state $\rho(\thetaopt)$ and the true ground state $ \sket{\phi_g}$ probes the quality of the VQE solution. A sudden drop in the overlap near the transition threshold indicates that the measured properties using the VQE solution no longer approximate the ground-state properties. This means VQE calculation can be extremely sensitive to small changes in error rates.

The observed transition can be understood by looking at the cost-function landscapes associated with the two-parameter ansatz before and after the transition shown in \cref{fig:schematic}(a). The two minima labeled minimum 1 and minimum 2 respond differently to the increase in the error rate. Minimum 1 is the global minimum for small $p_x$, but its value increases more rapidly with increasing $p_x$ compared to minimum 2. At the transition threshold and beyond, minimum 2 corresponding to a different ansatz state becomes the global minimum. The mechanism of the observed transition is similar to that of the first-order phase transition in Ginzburg–Landau theory.

The global minimum switching can be explained by the first-order perturbative correction of the cost-function landscape's response to noise. By expanding the ansatz density matrix $\rho(\vec{\theta}; p_x)$ in order of $p_x$ and retaining only the leading-order terms, we can write the noisy cost function as
\be
\label{eq:E_pert}
E(\vec{\theta}; p_x) = (1 - n_{\ell} N p_x ) \mathrm{Tr}(\rho_0(\vec{\theta}) H) +  p_x \mathrm{Tr}(\rho_1(\vec{\theta}) H),
\ee
where $\rho_0(\vec{\theta}) = \ketbra{\psi(\vec{\theta})}{\psi(\vec{\theta})}$ is the error-free ansatz state, $\rho_1(\vec{\theta}) \sim \mathcal{O}(n_{\ell} N )$ is the first-order correction and $n_{\ell}$ is the number of ansatz layer.
Optimizing this approximated cost function gives us the dashed lines shown in \cref{fig:dimer_result}(a), and the optimal solutions from the perturbed cost function are qualitatively consistent with that from the full cost function. This suggests that the first-order perturbative correction can account for the switching causing the transition, and the threshold error rate roughly scales as $1 / n_{\ell} N$. Nonetheless, the ansatz structure and $\rho_1(\vec{\theta})$ change accordingly when we vary $n_{\ell}$ or $N$. To the best of our knowledge, there is no way to extrapolate the threshold error rate from one ansatz to another.

The optimizer's solution distribution shown in \cref{fig:dimer_result}(c) is a helpful tool to inspect the cost-function landscape.
Directly inspecting the cost-function landscape for ansatzes with more than two parameters is difficult since we can only visualize the landscape with dimensionality reduction. We can use the optimizer's solutions obtained with different initial parameters (128 initial parameters per $p_x$) to overcome this challenge. The optimizer's solutions correspond to (local or global) minima close to the initial parameters since local optimizer is used. Hence, we can infer the energies associated with different minima by the distribution of the optimizer's solutions for the eight-parameter ansatz. The two sets of optimizer's solutions corresponding to minimum 1 (green dashed line) and minimum 2 (blue dashed line) cross each other, signaling the transition at around $p_x=0.07$. Note that after the transition, there is a third set of optimizer's solutions with energy between minima 1 and 2 irrelevant to the transition.

\section{Spin chains simulation}

\begin{figure}
	\centering
	\includegraphics[width=0.8\linewidth]{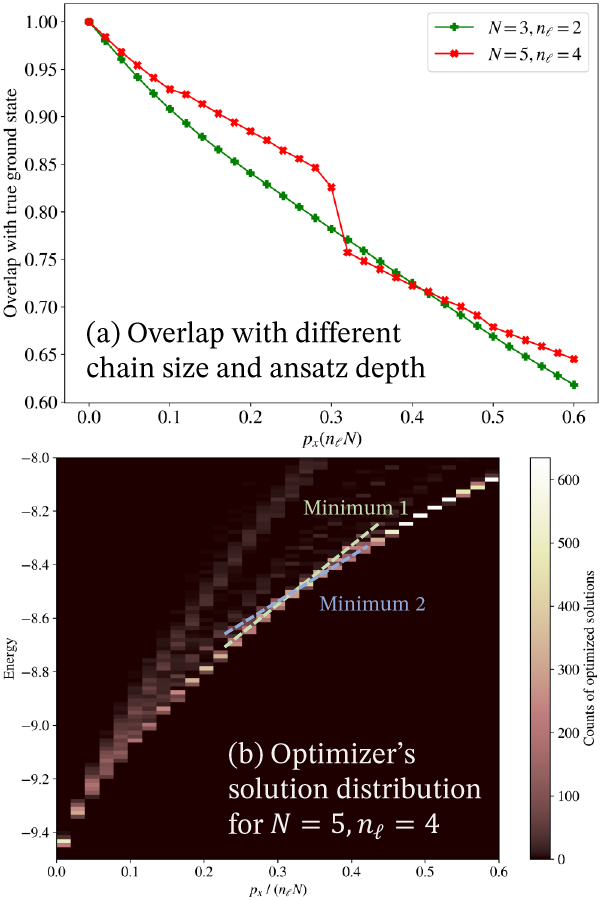}
	\caption{\textbf{Spin chain results of two system sizes and ansatz depths}.
	The optimal solution of VQE for a spin chain with system size $N=5$ and ansatz depth $n_{\ell}=4$ [red crosses in (a)] shows a transition at around $p_x n_{\ell} N = 0.3$. This transition corresponds to the crossover between two minima shown in the optimizer's solution distribution in (b). Even after increasing the number of random initial parameter sets from 128 to 768 to improve the figure resolution, the complicated landscape, due to a higher dimensional cost function (74 parameters), still makes it difficult to determine the crossover location from the distribution. The optimal solution with system size $N=3$ and ansatz depth $n_{\ell}=2$ [green pluses in (a)] shows no transition.
	} 
	\label{fig:chain_result}
\end{figure}

The noise-induced transition can also be observed in spin chains using the general ansatz shown in \cref{fig:schematic}(b). Similar to the dimer case, we employ the same optimization approach and plot the overlaps between the optimal solution and the true ground state against the error rate, as shown in \cref{fig:chain_result}(a), for two different system sizes $N$. The ansatz depth $n_{\ell}$ is chosen to be the minimum depth that can still accurately approximate the true ground state at $p_x=0$. We scale the blip-flip error rate by $n_{\ell} N$ according to the leading-order response to error in \cref{eq:E_pert}.

The chain with a length of $N=5$ exhibits a transition at approximately $p_x n_{\ell} N=0.3$ or $p_x = 0.015$.
This transition can be explained by the crossover of the two minima suggested by the optimizer's solution distribution shown in \cref{fig:chain_result}(b) with labeled minimum 1 and minimum 2. At around $p_x n_{\ell} N=0.1$, the distribution shows the emergence of several minima. The minima could be the reason for the sudden change in the slope of the overlap, and further study is required to determine the cause of this behavior.

The nontrivial effect of the ansatz's structure on the cost-function landscape makes extrapolating the existence of a transition difficult when varying $N$ and $n_{\ell}$. For example, while there are noise-induced transitions for $N=2, n_{\ell}=1$ and $N=5, n_{\ell}=4$ at around $p_x  n_{nll} N = 0.14$ and $p_x / n_{nll} N = 0.3$, respectively, we do not find any transition signature for $N=3, n_{ell}=2$ with similar error rates.
As there is no straightforward way to predict the transition, a close examination of data generated by variational algorithms is necessary, particularly when the data is collected from multiple QPUs or over an extended period of time in which hardware parameter drift is likely to occur.

\section{Dimer demonstration on IBM QPU}

\begin{figure}
	\centering
	\includegraphics[width=0.8\linewidth]{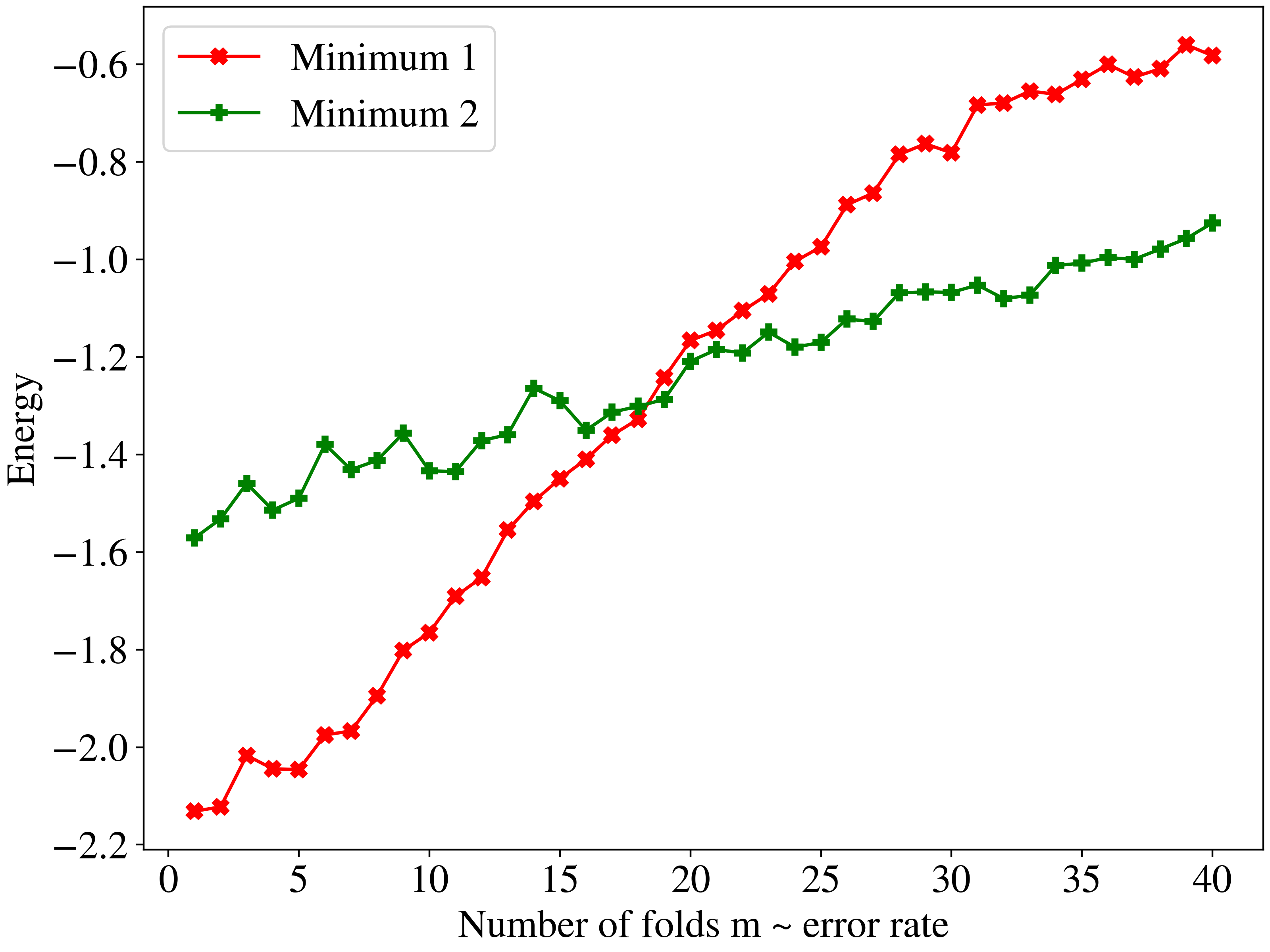}
	\caption{\textbf{Demonstration of the global minimum switching on IBM QPU}.
		For the spin dimer case using the two-parameter ansatz, the parameters corresponding to the two minima shown in \cref{fig:schematic}(a) are determined by classical simulation. The energies associated with these two minima are then measured on qubits 14 and 13 of ``ibm\_cairo" with an increasing number of $m$ approximating the increasing error rate. Even though the errors on the hardware cannot be accounted for by the simple bit-flip channel used in our numerical simulation, the global minimum switches from minimum 1 to minimum 2 at around $m=18$, indicating that a transition still occurs.
		We ran this demonstration on IBM Quantum through \emph{Qiskit} API \cite{Qiskit} on August 16, 2023. Each data point is measured with 8192 shots, and the result is mitigated by readout error mitigation.
	} 
	\label{fig:bim_result}
\end{figure}

We demonstrate the crossover of the two minima associated with the two-parameter ansatz for the spin dimer on the IBM QPU ``ibm\_cairo." While there is no direct control to tune the QPU error rates, we can increase the gate error rate by "folding" the CNOT gate in the same way as commonly used in digital zero-noise extrapolation \cite{Cai2023}. In particular, we fold the CNOT gate $m$ times with the following replacement rule,
\be
\mathrm{CNOT} \rightarrow \mathrm{CNOT} ( \mathrm{CNOT} \ \mathrm{CNOT} )^{m}.
\ee
The two CNOT gates inside the bracket reduce to an identity gate in the error-free limit, affecting the algorithm outcome only due to the associated errors. Hence, the number of folds $m$ can serve as an approximation of the error rate.
Moreover, it will be expensive to run the full optimization for different numbers of folds on the QPU. We instead only measure the cost function values associated with the two minima shown in \cref{fig:schematic}(a) with the minima's parameters determined classically.

The results presented in \cref{fig:bim_result} demonstrate that the global minimum switch from minimum 1 (red crosses) to minimum 2 (green pluses) at around $m=18$. There are various sources of error that contribute to the CNOT gate error, such as control error, qubit decoherence, and dephasing. This QPU demonstration that incorporates realistic noise suggests that the noise-induced transition is not specific to the bit-flip error channel we theoretically studied above.

\section{Conclusion}
In this manuscript, we show with numerical evidence and a QPU demonstration that the optimal solutions of noisy VQE algorithms can exhibit a transition behavior induced by an increasing error rate. This noise-induced transition is due to the global minimum of the cost function switching from one local minimum to another under the influence of noise, similar to the mechanism of first-order phase transitions. The transition could have a substantial negative impact on the solution's quality from a variational algorithm. It is also difficult to extrapolate whether the transition will occur from other ansatzes with a different depth or system size. Careful data analysis, for example, inspecting the local optimizer's solution distribution from NISQ hardware is necessary to avoid misinterpreting the noise-induced features as the simulated system's genuine properties.

\begin{acknowledgments}
This manuscript has been authored by Fermi Research Alliance, LLC under Contract No. DE-AC02-07CH11359 with the U.S. Department of Energy, Office of Science, Office of High Energy Physics.
A.C.Y.L. is partially supported by the DOE/HEP QuantISED program grant "HEP Machine Learning and Optimization Go Quantum", identification number 0000240323.
I.H. is supported by the Open Quantum Initiative undergraduate fellowship program.
This research used resources of the Oak Ridge Leadership Computing Facility, which is a DOE Office of Science User Facility supported under Contract DE-AC05-00OR22725.
We acknowledge the use of IBM Quantum services for this work. The views expressed are those of the authors, and do not reflect the official policy or position of IBM or the IBM Quantum team.
\end{acknowledgments}

\bibliographystyle{apsrev4-2}
\bibliography{bib}

\end{document}